\documentclass[%
reprint,
superscriptaddress,
 amsmath,amssymb,
 aps,
prl,
longbibliography,
floatfix,
]{revtex4-2}
\usepackage{graphicx}
\usepackage{color}
\usepackage{hyperref}
\usepackage{float}
\usepackage{array, multirow}
\usepackage{amsmath}
\usepackage{newtxtext}
\usepackage{booktabs}
\usepackage{xcolor}
\usepackage{soul}
\usepackage{url}
\usepackage[margin=0.8in]{geometry}

\hypersetup{colorlinks,
    linkcolor={blue!75!black!80!yellow},
    citecolor={blue!75!black!80!yellow},
    urlcolor={blue!75!black!80!yellow}
    }

\begin{document}

\title{Quantum Simulation of Open Quantum Systems Using Density-Matrix Purification}

\author{Anthony W. Schlimgen}
\affiliation{Department of Chemistry and The James Franck Institute, The University of Chicago, Chicago, IL 60637 USA}
\author{Kade Head-Marsden}
\affiliation{John A. Paulson School of Engineering and Applied Sciences, Harvard
University, Cambridge, MA 02138, USA}
\author{LeeAnn M. Sager-Smith}
\affiliation{Department of Chemistry and The James Franck Institute, The University of Chicago, Chicago, IL 60637 USA}
\author{Prineha Narang}
\affiliation{John A. Paulson School of Engineering and Applied Sciences, Harvard
University, Cambridge, MA 02138, USA}
\author{David A. Mazziotti}
\email{damazz@uchicago.edu}
\affiliation{Department of Chemistry and The James Franck Institute, The University of Chicago, Chicago, IL 60637 USA}

\date{Submitted July 14, 2022}

\begin{abstract}

Electronic structure and transport in realistically-sized systems often require an open quantum system (OQS) treatment, where the system is defined in the context of an environment.  As OQS evolution is non-unitary, implementation on quantum computers\textemdash limited to unitary operations\textemdash is challenging.  We present a general framework for OQSs where the system's $d \times d$ density matrix is recast as a $d^{2}$ wavefunction which can be evolved by unitary transformations.  This theory has two significant advantages over conventional approaches: (i) the wavefunction requires only an $n$-qubit, compared to $2n$-qubit, bath for an $n$-qubit system and (ii) the purification includes dynamics of any pure-state universe.  We demonstrate this method on a two-level system in a zero temperature amplitude damping channel and a two-site quantum Ising model. Quantum simulation and experimental-device results agree with classical calculations, showing promise in simulating non-unitary operations on NISQ quantum devices.
\end{abstract}

\maketitle

\noindent{\em Introduction:} Modeling quantum mechanical systems in realistic physical contexts frequently requires a careful partitioning into a system and an environment. Systems interacting with an environment, which are known as open quantum systems (OQS), are common in contemporary chemistry and physics, and several techniques to model the dynamics are available
~\cite{Breuer2002, deVega2017, Weimer2021}. Environmental interaction generally causes non-unitary evolution in the system, but the combined system-bath evolution remains unitary. With the increased use of quantum algorithms for unitary-gate-based quantum computers, techniques for casting non-unitary operations into unitary forms have become increasingly important.

Increasing the size of the system or embedding the system in an effective environment, known generally as a dilation, allows for the modeling of the reduced dynamics of the system, without a complete description of the bath degrees of freedom. A critical question is the necessary size of the effective bath to represent the dynamics of the system. In a seminal study on quantum computing Lloyd conjectured that the bath need only be of dimension $d$, which is the size of the system density matrix~\cite{Lloyd1996}; however, this conjecture was putatively disproven a short time later when the analysis of quantum channels suggested the bath size is bounded from above by $d^2$~\cite{Terhal1999}.  A second important question, which we relate to the first question, is the range of physical dynamics that should be included within an OQS theory.  The conventional approach to OQS, based on quantum channels, assumes that the initial density matrix is separable in its system and bath---an assumption that has been recognized to fail in many realistic OQS scenarios such as electronic systems and environments or non-Markovian dynamics~\cite{Pechukas1994, Alicki1995}.  

In this {\em Letter} we show that a general OQS can be modeled with unitary dynamics by embedding the system in a bath with size exactly $d$. This is achieved through a process known as density-matrix purification~\cite{Hughston1993, Schumacher1996, Bassi2003, Kleinmann2006, Umezawa1996}, where a mixed-state density matrix is represented by a pure-state wavefunction of length $d^2$. Purification techniques have been used elsewhere in particular in thermal-state electronic structure~\cite{Knizia2013,BenavidesRiveros2022,Miller2019,Ke:2022wn,Borrelli:2019va,Borrelli:2021uw,Verstraete:2004ut,Feiguin:2005wl}, quantum information~\cite{Nielsen2010}, and quantum gravity~\cite{Wu:2019wx,Zhu:2020uv}. Considering open quantum system dynamics through density-matrix purification has several advantages over conventional quantum channel approaches. First, because the size of the bath is bounded from above by $d$ in purification, it is more computationally efficient than the quantum channels approach in the context of both classical and quantum algorithms.  This is particularly important in near-intermediate-scale quantum (NISQ) computing where the reduction in the bath size results in savings in both qubits and entangling gates. Second, the purification approach achieves its greater efficiency by broadening the class of allowed OQS dynamics treatable by conventional approaches based on quantum channels~\cite{Pechukas1994, Alicki1995, Shaji2005, Dominy2016}, providing a general framework for analyzing OQS dynamics when the system and bath are initially entangled, or when the dynamical evolution is not completely positive.  Finally, the density-matrix purification, in contrast to dilations based on a single qubit~\cite{Hu2020, HeadMarsden2021, Schlimgen2021, Hu2022, SchlimgenSVDArxiv}, allows for an arbitrary number of non-unitary transformations to be performed sequentially on a quantum device without any reliance upon classical devices for intermediary storage or computations.

After developing the purification theory, we demonstrate the method's promise for non-unitary simulations on NISQ devices through quantum simulation and experimental-device results for a two-level system in a zero temperature amplitude damping channel and a quantum Ising model. 

\noindent {\em Theory:} Consider an open quantum system consisting of a quantum system and a bath.  The reduced density matrix of the system at time $t$ can be expressed in terms of its spectral expansion
\begin{equation}\label{eq:dm}
\rho_{\rm S}(t) = \sum_{k}{ w_{k}(t) | \Psi_{k}^{\rm S}(t) \rangle \langle \Psi_{k}^{\rm S}(t) | },
\end{equation}
where $w_{k}(t)$ are the eigenvalues and $| \Psi_{k}^{\rm S}(t) \rangle$ are the eigenfunctions. In a process known as purification, the ensemble of the system can be embedded into a pure-state wavefunction of the system and an effective bath
\begin{equation}\label{eq:psi}
| \Psi(t) \rangle = \sum_{k}{ \sqrt{w_{k}(t)} | \Psi_{k}^{\rm S}(t) \rangle | \Psi_{k}^{\rm B}(t) \rangle },
\end{equation}
in which the dimension of the effective bath is the same dimension of the system~\cite{Hughston1993, Schumacher1996, Bassi2003, Kleinmann2006, Knizia2013, MullerHermes2018, BenavidesRiveros2022, Schmidt1906}. Taking the trace over the bath degrees of freedom of the system-bath density matrix yields the reduced density matrix of the system
\begin{equation}\label{eq:tr}
\rho_{\rm S}(t) = {\rm Tr}_{\rm B} \left ( | \Psi^{\rm SB}(t) \rangle \langle \Psi^{\rm SB}(t) |  \right ) .
\end{equation}
A larger bath can always be represented by a bath of the system dimension because the system of dimension $d$ can only interact with a maximum of $d$ degrees of freedom in the bath.  The decomposition of the wavefunction in Eq.~(\ref{eq:psi}) is often known as the Schmidt polar form~\cite{Hughston1993, Schmidt1906}.  We have demonstrated the first necessary result for an open quantum system based on purification: {\em any system density matrix can be represented by a pure-state system-bath wavefunction in which the system and the bath share the same dimensions.}

The time evolution of the system-bath wavefunction can be generally expressed by a system-bath unitary transformation
\begin{equation}\label{eq:psit}
| \Psi^{\rm SB}(t') \rangle = {\hat U}_{\rm SB}(t';t) | \Psi^{\rm SB}(t) \rangle.
\end{equation}
Similarly, the system reduced density matrix at time $t'$ can be obtained from inserting Eq.~(\ref{eq:psit}) into Eq.~(\ref{eq:tr}).  The general unitary transformation operator can be written as follows
\begin{equation}\label{eq:u}
{\hat U}_{\rm SB}(t';t) = {\rm e}^{i {\hat H}_{\rm S}(t';t) \otimes I + I \otimes i {\hat H}_{\rm B}(t';t)+i {\hat H}_{\rm SB}(t';t)},
\end{equation}
in which ${\hat H}_{\rm S}(t';t)$, ${\hat H}_{\rm B}(t';t)$, and ${\hat H}_{\rm SB}(t';t)$ are Hermitian operators on the system, the bath, and the system-bath, respectively.  The entanglement of the system and the bath, controlled by the unitary transformation, kindles the formation of an ensemble density matrix on the system.  Because any system-bath wavefunction at time $t'$ can be reached by unitary transformation, we can generate any system reduced density matrix at time $t'$. Consequently, we have demonstrated the second necessary result for treating open quantum systems by purification: {\em any system reduced density matrix can be generated by a unitary transformation that mixes the system and bath degrees of freedom of its purification wavefunction in which both the system and the bath share the same dimension $d$.}

In combination, these two results allow us to establish a theory of open quantum systems based on density-matrix purification. While the quantum channels are represented by Kraus or Lindblad operators in conventional approaches to open quantum systems~\cite{Breuer2002, Lindblad1976}, in the present approach the mixing of the system and the bath by the unitary transformation modulates the degree of the entanglement between the system and its bath. As the entanglement increases, the system density matrix becomes more ensemble (mixed state), and as the entanglement decreases, the system density matrix becomes increasingly pure.  Importantly, the selection of the unitary transformation determines the time evolution of the system density matrix and its effective interaction with the bath.  While the bath density matrix shares the eigenvalues of the system density matrix, the eigenfunctions of the bath density matrix are arbitrary.  For concreteness, we can select the bath eigenfunctions to be identical to the eigenfunctions of the system density matrix.

The theory of open quantum systems based on density-matrix purification has several key attributes. First, the evolution of the open quantum system is performed by a unitary transformation of the wavefunction from the purification of the system density matrix.  Such a formulation is particularly important for applications on quantum computers where the unitary transformation described here can be implemented with $2n$ qubits where $n$ ($=\log_{2}(d)$) is the number of qubits required to describe the $d$-dimensional system. Second, the density-matrix purification includes all density matrices that are expressible as a reduced density matrix of a pure state. Such density matrices are critically important to describing electronic density matrices that are part of a larger molecule.

The density-matrix purification approach to open quantum systems can be compared with the conventional quantum-channel approach. In the conventional approach the evolution of the system density matrix is described by a set of Kraus maps with each Kraus operator $M_{k}$ representing a quantum channel
\begin{equation}\label{eq:kraus}
\rho_{\rm S}(t') = \sum_{k}{{\hat M}_{k} \rho_{\rm S}(t) {\hat M}_{k}^{\dagger} } .
\end{equation}
The Kraus map approach is equivalent to the following evolution of the system density matrix by unitary transformation of the system and bath density matrices~\cite{Breuer2002, Schumacher1996}
\begin{align}\label{eq:stin}
\rho_{\rm S}(t') = & {\rm Tr}_{\tilde {\rm B}} \left ( \hat U_{{\rm S}{\tilde {\rm B}}} \left ( \rho_{\rm S}(t) \otimes | \Psi_{0}^{\tilde {\rm B}} \rangle \langle \Psi_{0}^{\tilde {\rm B}} | \right ) \hat U_{{\rm S}{\tilde {\rm B}}}^{\dagger} \right ) \\
                 = & \sum_{k}{ w_{k} {\rm Tr}_{\tilde {\rm B}} \left ( \hat U_{{\rm S}{\tilde {\rm B}}} | \Psi_{k}^{\rm S} \rangle | \Psi_{0}^{\tilde {\rm B}} \rangle \langle \Psi_{0}^{\tilde {\rm B}} | \langle \Psi_{k}^{{\rm S}} |  \hat U_{{\rm S}{\tilde {\rm B}}}^{\dagger} \right ) },
\end{align}
where in the second line we inserted the spectral expansion of the system density matrix. Importantly, this expression for the system density matrix is significantly different from our expression obtained in the context of density-matrix purification
\begin{equation}\label{eq:pure}
\rho_{\rm S}(t') = \sum_{kl}{ \gamma_{kl} {\rm Tr}_{\rm B} \left ( {\hat U}_{\rm SB} | \Psi_{k}^{\rm S} \rangle | \Psi_{k}^{\rm B} \rangle \langle \Psi_{l}^{\rm S} | \langle \Psi_{l}^{\rm B} | {\hat U}_{\rm SB}^{\dagger} \right ) },
\end{equation}
where $\gamma_{kl} = \sqrt{w_{k}(t) w_{l}(t)}$.  While the system and bath are not entangled at the initial time in the conventional description of open quantum systems, the system and bath are significantly entangled in the density-matrix purification prior to the action of the unitary matrix.

The density-matrix purification theory for open quantum systems has two major advantages over the conventional theory.  First, the dimension of the bath ${\tilde {\rm B}}$ in the conventional formulation must be $d^{2}$ to be complete~\cite{Terhal1999, Lloyd1996} while the dimension of the bath ${\rm B}$ in the purification form is only $d$ to be complete.  Hence, the density-matrix purification is computationally more efficient than the Kraus maps formalism. In the context of quantum computing, the computation of open quantum systems by purification requires $2n$ qubits---$n$ qubits for the system and $n$ qubits for the bath, but computation by Kraus maps requires potentially $3n$ qubits---$n$ qubits for the system and $2n$ qubits for the bath.

Second, the set of density matrices in the conventional formulation is a subset of the set of density matrices in the purification approach. By construction, its density-matrix purification covers all density matrices that arise from a system that is a part of a larger pure-state universe (system plus bath). Consequently, the conventional approach, even with its greater computational cost, does not capture important open systems that arise from a system embedded in a larger setting, such as a sub-molecule in the context of a larger molecule. Formally, the density matrices in the conventional approach are said to be completely positive while the density matrices in the purification approach are positive, that is containing all trace-preserving, Hermitian, and positive-semidefinite density matrices. While complete positivity has been traditionally viewed as an advantage, here, as seen elsewhere, we observe that the completely positive set omits a significant family of physical density matrices~\cite{Pechukas1994, Alicki1995, Shaji2005,  Dominy2016, Schmid2019}.

\noindent {\em Methods:} To explore the OQS theory based on density-matrix purification, we compute for two well-known models the unitary transformations in Eq.~(\ref{eq:u}) that match the dynamics from a quantum-channels-based theory.  For each system, we generate the unitary transformations through three steps: (i) computation of the exact dynamics using Lindblad's equation, (ii) determination of the purified system-bath wavefunctions $| \Psi^{SB}(t^\prime) \rangle$, and finally, (iii) solution of the unitary transformations $U_{SB}(t',t)$ that yields the purified wavefunctions.

First, we generate the exact system dynamics using the unravelled Lindblad equation
\begin{equation}
    \lvert \rho_S(t') \rangle = e^{\hat{\mathcal{L}}t'} \lvert \rho_S(0) \rangle,
    \label{eq:prop}
\end{equation}
where $\lvert \rho_S(t') \rangle$ is the vectorized density matrix at time $t'$, and the Lindbladian superoperator, $\hat{\mathcal{L}}$ is given by
\begin{equation}
\begin{aligned}
    \hat{\mathcal{L}} &= -i\mathbb{I}\otimes \hat{H} + i\hat{H}^T\otimes \mathbb{I}
    \\
    &+ \sum_k \hat{C}^*_k \otimes \hat{C}_k -\frac{1}{2}\mathbb{I}\otimes(\hat{C}_k^{\dagger}\hat{C}_k)
    - \frac{1}{2}\hat{C}_k^T\hat{C}^*_k\otimes \mathbb{I},
\end{aligned}
\label{eq:lindblad}
\end{equation}
in which $\hat H$ is the Hamiltonian and the $\hat C_i$'s are the dissipative Lindbladian terms~\cite{Lindblad1976, Gorini1976, Breuer2002}. The Kraus operators and Eq.~(\ref{eq:prop}) generate the same dynamics; however, the Kraus operators are not always known. Further discussion of the unravelled Lindblad equation and Kraus maps is available elsewhere~\cite{Havel2003, Schlimgen:2022ul}.

Second, we optimize $| \Psi^{SB}(t^\prime) \rangle$ by minimizing the objective
\begin{equation}
\begin{aligned}
\textrm{min} \quad &\| \rho_{\rm S}(t') - {\rm Tr}_{\rm B} (\rho_{\rm SB}(t') ) \| , \\
\end{aligned}
\end{equation}
where
\begin{equation}
    \rho_{\rm SB}(t') = | \Psi^{SB}(t^\prime) \rangle \langle \Psi^{SB}(t^\prime) |,
    \label{eq:sbDM}
\end{equation}
the system and bath density matrices are constrained to be equal, and $\| \cdot \|$ is the Frobenius norm. This objective ensures that the purified, optimized density matrix contracts to the exact system density matrix whether contraction is performed with respect to either the system or bath degrees of freedom.

Finally, we generate each unitary transformation $U_{SB}(t',t)$ by optimizing $\hat{H}_{SB}(t^\prime,t)$ 
\begin{equation}
      \textrm{min} \quad \| | \Psi^{SB}(t^\prime) \rangle - U_{SB}(t',0) | \Psi^{SB}(0) \rangle \|.
    \label{eq:minU} 
\end{equation}
A general complex Hermitian operator will have $d^2$ independent variables, where $d$ is the dimension of the matrix corresponding to $\hat H_{\rm SB}(t';t)$.
We perform all optimizations using the Broyden–Fletcher–Goldfarb–Shanno (BFGS) quasi-Newton algorithm implemented in the Python package SciPy~\cite{bfgs2006}.

\begin{figure}
    \centering
    \includegraphics[width=\columnwidth]{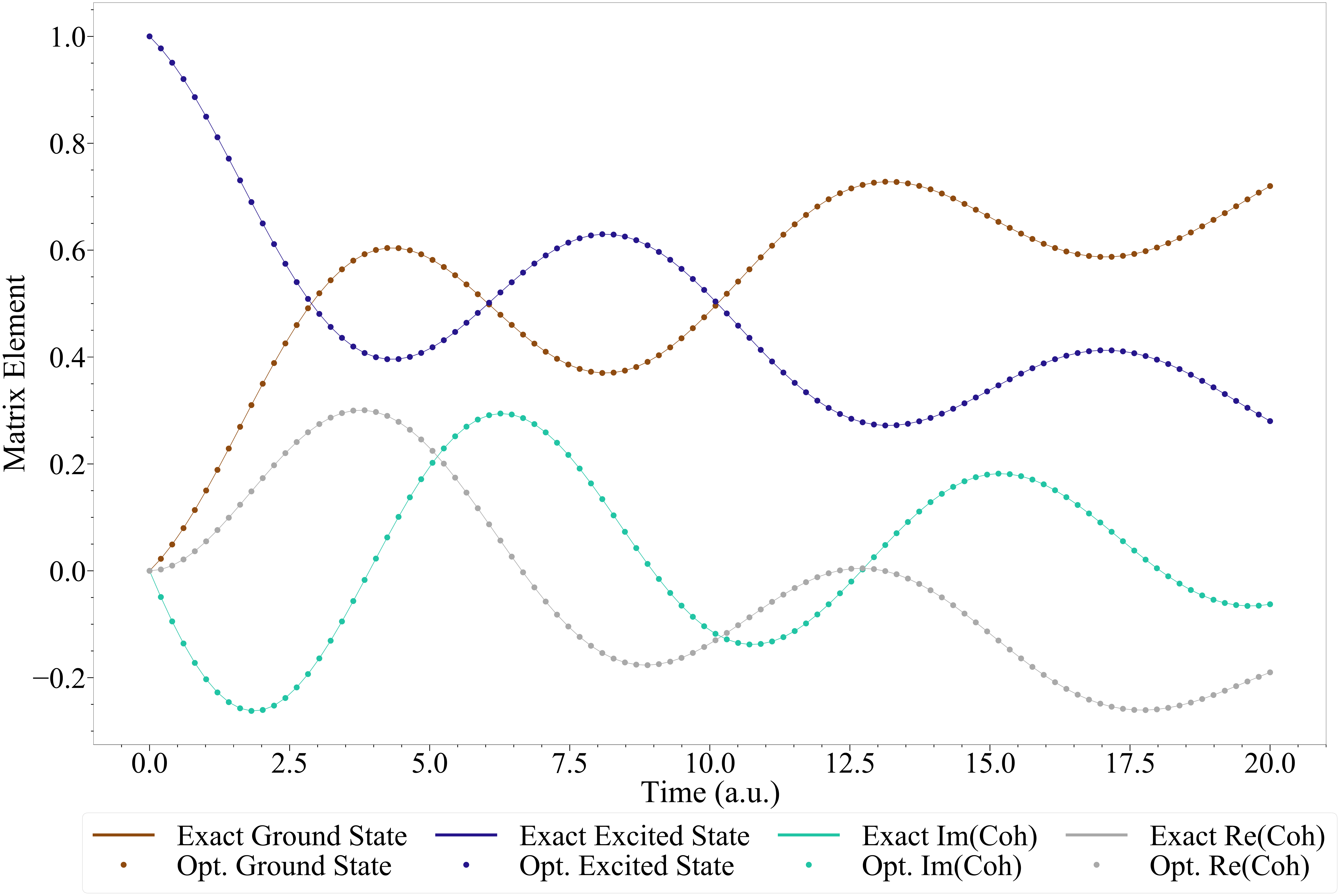}
    \caption{Classical simulation of the two-level driven amplitude damping channel from density-matrix purification. The exact results are solid lines and the results from optimization and unitary propagation are dots. The optimized results are in excellent agreement with the exact solution.}
    \label{fig:tls_class}
\end{figure}

\noindent {\em Results:} We show the dynamics for two open quantum systems using the density-matrix purification approach. First, we show the dynamics of a two-level system (TLS) in an amplitude damping channel at zero temperature. These dynamics are generated on a classical computer. The system Hamiltonian is
\begin{equation}
\hat{H} = -\frac{\delta}{2}\hat{\sigma}_z - \frac{\Omega}{2}\hat{\sigma}_x,
\end{equation}
where $\delta$ is the detuning parameter, and $\Omega$ is the Rabi frequency. The Kraus maps for this channel are
\begin{equation}
\begin{aligned}
    M_0 &= \begin{pmatrix}
            1 & 0\\
            0 & \sqrt{e^{-\gamma t}}
          \end{pmatrix}\\
    M_1 &= \begin{pmatrix}
            0 & \sqrt{1-e^{-\gamma t}}\\
            0 & 0
          \end{pmatrix},\\
          \end{aligned}
          \label{eq:gen_amp}
\end{equation}
where $\gamma$ is the damping strength. The corresponding Lindbladian operator is $\hat{C}_0=\sqrt{\gamma}\hat \sigma_-$, where $\hat \sigma_-$ is the lowering operator. The dynamics of this open system are shown in Fig.~\ref{fig:tls_class} using the parameters $\Omega=\delta=0.5$, and $\gamma=0.1$, where the exact solution is the solid lines, and the optimized solution is the dots. The system is initialized in the excited state, and energy dissipates over time as the ground state is repopulated. The average distance between the exact and optimized system density matrices is $\mathcal{O}(10^{-6})$ over the time frame, indicating excellent agreement between the Lindblad solution and the purification solution.

\begin{figure}
    \centering
    \includegraphics[width=\columnwidth]{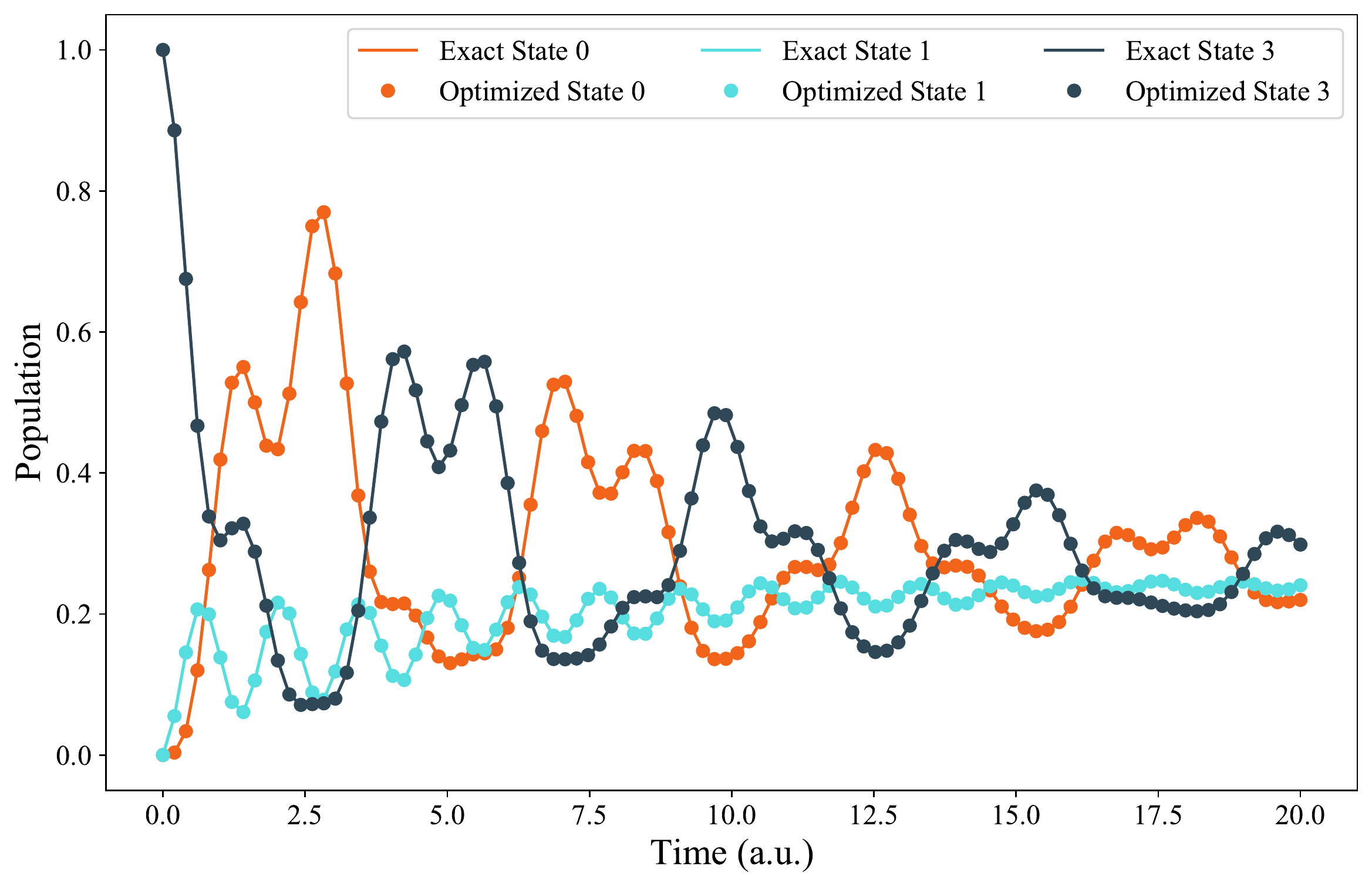}
    \caption{Classical solution of the two-site transverse-field Ising model optimized for density-matrix purification. The exact solutions are the solid lines and the optimized solution is the dots. There is excellent agreement between the exact solution and the optimized solution for density-matrix purification.}
    \label{fig:tfim_class}
\end{figure}

A second example is simulating, also classically, the damped two-site transverse-field Ising model (TFIM), which is a four-level system. The Hamiltonian for this system is
\begin{equation}
    \hat{H} = J\sum_i \hat{\sigma}_z^i \hat{\sigma}_z^{i+1} - h \sum_i \hat{\sigma}_x^i,
\end{equation}
where $J$ is the nearest-neighbor coupling strength and $h$ is the transverse magnetic field, and the dissipation is dictated by the Lindbladian for each channel $\hat{C}_i = \sqrt{\gamma}\hat{\sigma}^i_-$, where $\gamma$ is the damping parameter. The dynamics for the two-site TFIM are shown in Fig.~\ref{fig:tfim_class}, with both sites initially in the excited state, $J=h=1$, and $\gamma=0.1$. Similarly to the TLS, the TFIM dynamics via density-matrix purification results in excellent agreement with the exact dynamics from the Lindbladian approach. The average error in the optimized density matrix is again $\mathcal{O}(10^{-6})$.

\begin{figure}
    \centering
    \includegraphics[width=\columnwidth]{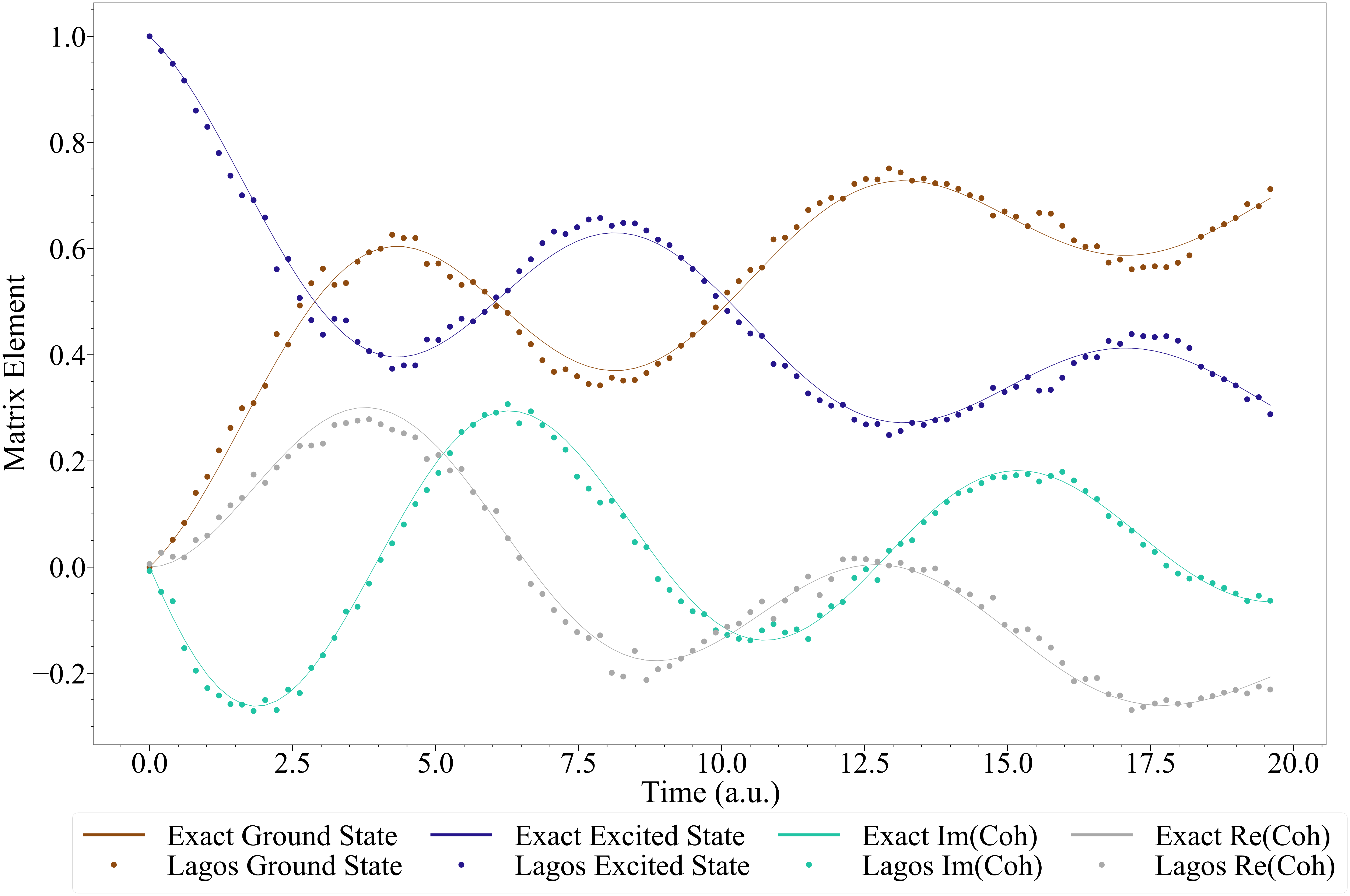}
    \caption{Results of the TLS simulated with IMBQ Lagos quantum computer, with exact results as solid lines and results from the quantum device as dots. Both the populations and coherences show good agreement with the exact solution. Results generated using 32000 shots and Qiskit error mitigation.}
    \label{fig:tls_lagos}
\end{figure}

We also perform the evolution of the TLS by quantum simulation. We perform the propagation on the IBMQ Lagos quantum device, with 32000 shots (samples)~\cite{lagos}. We used Qiskit's gate transpiler to implement the unitaries in basic quantum gates, as well as built-in error mitigation tools~\cite{Qiskit}. We use the unitaries generated from the classical optimization.  The results from the quantum device are shown in Fig.~\ref{fig:tls_lagos}.
The simulation requires a circuit with 2 qubits to model the purified state. We perform full tomography to reconstruct all elements of the density matrix, which requires 3 circuits per time point that can be run in parallel. Over the simulation time frame, the average error in the simulated density matrix is $0.04 \pm 0.02$.

\noindent {\em Conclusions:} 
Traditional approaches to open quantum systems have provided several avenues for understanding the dynamics of these systems, but persistent questions have stymied application of these techniques in certain contexts. The theory of OQS through density-matrix purification clarifies several issues, resulting in a more efficient and more broadly applicable theory. Here we have shown that the size of the dilated system is bounded by the size of the original open system, $d$, whereas in the theory of quantum channels, the bound is $d^2$. While here we optimize the unitary transformations mixing the system and bath to match known quantum channels, these transformations can be directly chosen within the OQS density-matrix purification to model the system-environment interactions in lieu of traditional quantum channels in the form of Kraus maps or Lindblad operators.  Furthermore, the purification approach presented here is applicable to a broader range of physically important problems than quantum channels, in particular, when the initial state is entangled with the bath, or in the context of embedded electronic systems.

This purification approach is more cost-effective in both classical and quantum computational contexts. While the classical savings is notable, the savings is particularly important when studying OQS's with NISQ devices, because the larger dilation requires more qubits and generally results in deeper circuits.  Furthermore, in comparison to single- or few-qubit dilations, non-unitary transformations can be performed repeatedly on the quantum device by density-matrix purification without any intermediate measurements, classical processing, or any decay of the probability amplitudes from unphysical outcomes.  We benchmarked this method on a two-level system in a zero temperature amplitude damping channel and the dissipative transverse field Ising model, showing results in excellent agreement with the classical channels approach. Moreover, we simulated the two-level system on IBM's Lagos quantum computer, demonstrating the application of the method to current quantum hardware.  The density-matrix purification theory provides a broad paradigm for treating open quantum systems with greater flexibility and efficiency than conventional approaches, and has important applications on quantum computers to modeling open quantum systems as well as more general non-unitary transformations.

\noindent {\em Acknowledgments}
This work is supported by the NSF RAISE-QAC-QSA, Grant No. DMR-2037783 and the Department of Energy, Office of Basic Energy Sciences Grant DE-SC0019215. D.A.M. also acknowledges NSF EAGER-QAC-QSA, Grant No. CHE-2035876.  We acknowledge the use of IBM Quantum services for this work. The views expressed are those of the authors, and do not reflect the official policy or position of IBM or the IBM Quantum team.

\bibliography{main}

\end{document}